\documentclass[a4paper,10pt,twocolumn,preprint,5p,lefttitle,nonatbib]{elsarticle}

\pdfoutput=1

\usepackage[table,xcdraw,svgnames,dvipsnames]{xcolor}
\usepackage[colorlinks,pdfencoding=auto]{hyperref}

\AtBeginDocument{%
  \hypersetup{
    citecolor=[rgb]{0.0941,0.50196,0.67451}, %DeepSkyBlue,
    linkcolor=[rgb]{0.0941,0.50196,0.67451},   
    urlcolor=[rgb]{0.0941,0.50196,0.67451},
    pdfauthor=author}
}
\usepackage[utf8]{inputenc}
\usepackage{booktabs}
\usepackage{url}
\usepackage{color}
\usepackage{multirow}
\usepackage{amsmath,amssymb,amsfonts}
\usepackage{graphicx}
\usepackage{textcomp}
\usepackage{listings}
\usepackage{soul}
\sethlcolor{white}

\makeatletter
\let\c@author\relax
\makeatother

\usepackage[backend=biber,style=authoryear,sortcites=none,minnames=3,maxnames=3,mincitenames=1,maxcitenames=2]{biblatex}
\usepackage[noabbrev,capitalize]{cleveref}

\usepackage{dsfont} % for math 1 symbol
\usepackage{color,array}

\usepackage{comment}

\usepackage{threeparttable} % convenient for having "captions" both above and underneath table

\usepackage{colortbl}

\usepackage{adjustbox}  % for resizing tables to fit page
\usepackage{afterpage}  % force figure to be set at next page (under constraints)

\usepackage{enumerate}% http://ctan.org/pkg/enumerate

\usepackage{diagbox}
\usepackage{nidanfloat} % for table across two coloumns at the bottom

\makeatletter
\newcommand*{\rom}[1]{\expandafter\@slowromancap\romannumeral #1@}
\makeatother

\usepackage{pifont}
\makeatletter
\newcommand{\customcref}[2]{%
   \protected@write \@auxout {}{\string \newlabel {#1}{{#2}{\thepage}{#2}{#1}{}} }%
   \hypertarget{#1}{#2}
}
\makeatother
\DefineBibliographyStrings{english}{%
    andothers = {\em et\addabbrvspace al\adddot}
}

\makeatletter
\def\printFirstPageNotes{%
  \iflongmktitle
   \let\columnwidth=\textwidth\fi
  \ifx\@tnotes\@empty\else\@tnotes\fi
  \ifx\@nonumnotes\@empty\else\@nonumnotes\fi
  \ifx\@cornotes\@empty\else\@cornotes\fi
  \ifx\@elseads\@empty\relax\else
   \let\thefootnote\relax
   \footnotetext{\ifnum\theead=1\relax
      \textit{E-mail address:\space}\else
      \textit{E-mail addresses:\space}\fi
     \@elseads}\fi
  \ifx\@elsuads\@empty\relax\else
   \let\thefootnote\relax
   \footnotetext{\textit{URL:\space}%
     \@elsuads}\fi
  \ifx\@fnotes\@empty\else\@fnotes\fi
  \iflongmktitle\if@twocolumn
   \let\columnwidth=\Columnwidth\fi\fi
}
\makeatother

\let\oldthebibliography\thebibliography
\renewcommand{\thebibliography}[1]{%
  \oldthebibliography{#1}
  \let\oldbibitem\bibitem
  \let\oldtextsc\textsc
  \def\oldbbland{et}
  \newcounter{authorcount}
  \def\bibitem[##1]##2{%
    \let\textsc\oldtextsc
    \let\bbland\oldbbland
    \oldbibitem[##1]{##2}%
    \let\textsc\mytextsc%
    \let\bbland\mybbland
    \setcounter{authorcount}{0}
  }
  \def\mybbland{\setcounter{authorcount}{0}\oldbbland}
  \def\dropetal##1.{ \bbletal}
  \def\mytextsc##1{%
    \oldtextsc{##1}%
    \stepcounter{authorcount}%
    \ifnum\value{authorcount}=5\relax%
      \expandafter\dropetal%
    \fi%
  }%
}

\makeatletter
\newbibmacro*{cite:plabelyear+extradate}{%
  \iffieldundef{labelyear}{}
    {\clearfield{labelmonth}% don't want months in citations
     \clearfield{labelday}% don't want days in citations
     \clearfield{labelendmonth}% don't want months in citations
     \clearfield{labelendday}% don't want days in citations
     \iffieldsequal{labelyear}{labelendyear}% Don't want no-op year ranges
       {\clearfield{labelendyear}}
       {}%
     \iffieldundef{origyear}
       {}
       {\printorigdate%
        \setunit*{\addslash}}%
     \iffieldundef{related}
       {}
       {\iffieldequalstr{relatedtype}{reprintfrom}
         {\entrydata*{\thefield{related}}{\printlabeldateextra}%
          \setunit*{\addslash}}
         {}}%
     \printlabeldateextra}}

\renewbibmacro*{cite}{%
  \iffieldequals{fullhash}{\cbx@lasthash}
   {\setunit{\compcitedelim}%
    \printtext[bibhyperref]{%
      \usebibmacro{cite:plabelyear+extradate}}}%
   {\printtext[bibhyperref]{%
      \ifnameundef{labelname}
       {\usebibmacro{cite:noname}%
         \setunit{\printdelim{nameyeardelim}}%
         \usebibmacro{cite:plabelyear+extradate}%
         \savefield{fullhash}{\cbx@lasthash}}
       {\ifnameundef{shortauthor}
         {\printnames{labelname}}%
         {\cbx@apa@ifnamesaved
           {\printnames{shortauthor}}
           {\ifnameundef{groupauthor}
             {\printnames[labelname]{author}}
             {\printnames[labelname]{groupauthor}}%
            \addspace\printnames[sabrackets]{shortauthor}}}%
         \setunit{\printdelim{nameyeardelim}}%
        \usebibmacro{cite:plabelyear+extradate}%
        \savefield{fullhash}{\cbx@lasthash}}}}%
   \setunit{\multicitedelim}}

\renewbibmacro*{textcite}{%
  \ifnameundef{labelname}
    {\iffieldundef{shorthand}
       {\printtext[bibhyperref]{%
          \usebibmacro{cite:label}}%
        \setunit{%
          \global\booltrue{cbx:parens}%
          \printdelim{nonameyeardelim}\bibopenparen}%
        \ifnumequal{\value{citecount}}{1}
          {\usebibmacro{prenote}}
          {}%
        \printtext[bibhyperref]{\usebibmacro{cite:labeldate+extradate}}}
       {\printtext[bibhyperref]{\usebibmacro{cite:shorthand}}}}
    {\printtext[bibhyperref]{\printnames{labelname}}%
     \setunit{%
       \global\booltrue{cbx:parens}%
       \printdelim{nameyeardelim}\bibopenparen}%
     \ifnumequal{\value{citecount}}{1}
       {\usebibmacro{prenote}}
       {}%
     \usebibmacro{citeyear}}}

\makeatother
\usepackage{fancyhdr}
\makeatletter
\let\runauthor\@author
\let\runtitle\@title
\makeatother
\newcommand{\abbreviations}[1]{%
  \nonumnote{\textbf{Abbreviations:\enspace}#1}}

\title{\textbf{Hybrid guiding: A multi-resolution refinement approach for semantic segmentation of gigapixel histopathological images}}

\begin{document}

\author[1,2]{Andr\'e Pedersen\corref{cor1}}
\ead{andre.pedersen@ntnu.no}
\author[3,4]{Erik Smistad}
%\author[1]{Anna M. Bofin}
\author[1,5]{Tor V. Rise}
\author[1,5]{Vibeke G. Dale}
\author[1,5]{Henrik S. Pettersen}
\author[6]{Tor-Arne S. Nordmo}
\author[4]{David Bouget}
\author[3,4]{Ingerid Reinertsen}
\author[1,2,5,7]{Marit Valla}

\address[1]{Department of Clinical and Molecular Medicine, Norwegian University of Science and Technology, NO-7491 Trondheim, Norway}
\address[2]{Clinic of Surgery, St. Olavs Hospital, Trondheim University Hospital, NO-7030 Trondheim, Norway}
\address[3]{Department of Circulation and Medical Imaging, Norwegian University of Science and Technology, NO-7491 Trondheim, Norway}
\address[4]{Department of Health Research, SINTEF Digital, NO-7465 Trondheim, Norway}
\address[5]{Department of Pathology, St. Olavs Hospital, Trondheim University Hospital, NO-7030 Trondheim, Norway}
\address[6]{Department of Computer Science, UiT The Arctic University of Norway, NO-9019 Tromsø, Norway}
\address[7]{Clinic of Laboratory Medicine, St. Olavs hospital, Trondheim University Hospital, NO-7030 Trondheim, Norway}

\cortext[cor1]{Corresponding author}

\abbreviations{WSI - whole slide image, CNN - convolutional neural network, PCA - principal component analysis, CAE - convolutional autoencoder, GUI - graphical user interface, GPU - graphics processing unit, CPU - central processing unit, Mob - MobileNetV2, Inc - InceptionV3, AN - annotation. BCS - Breast Cancer Subtypes. H\&E - hematoxylin-eosin, PW - patch-wise, LR - low-resolution, DSC - dice similarity coefficient, MIL - multiple instance learning.}

\begin{abstract}
Over the past decades, histopathological cancer diagnostics has become more complex, and the increasing number of biopsies is a challenge for most pathology laboratories. 
%Breast cancer tissue is heterogeneous and known for high inter-observer variability. 
Thus, development of automatic methods for evaluation of histopathological cancer sections would be of value.

In this study, we used 624 whole slide images (WSIs) of breast cancer from a Norwegian cohort.
We propose a cascaded convolutional neural network design, called H2G-Net, for semantic segmentation of gigapixel histopathological images. The design involves a detection stage using a patch-wise method, and a refinement stage using a convolutional autoencoder. To validate the design, we conducted an ablation study to assess the impact of selected components in the pipeline on tumour segmentation.

Guiding segmentation, using hierarchical sampling and deep heatmap refinement, proved to be beneficial when segmenting the histopathological images.
We found a significant improvement when using a refinement network for post-processing the generated tumour segmentation heatmaps.
The overall best design achieved a Dice score of $0.933 \pm 0.069$ on an independent test set of 90 WSIs. The design outperformed single-resolution approaches, such as cluster-guided, patch-wise high-resolution classification using MobileNetV2 ($0.872 \pm 0.092$) and a low-resolution U-Net ($0.874 \pm 0.128$). In addition, the design performed consistently on WSIs across all histological grades and segmentation on a representative $\times$400 WSI took $\sim$ 58 seconds, using only the central processing unit.

The findings demonstrate the potential of utilizing a refinement network to improve patch-wise predictions. The solution is efficient and does not require overlapping patch inference or ensembling.
Furthermore, we showed that deep neural networks can be trained using a random sampling scheme that balances on multiple different labels simultaneously, without the need of storing patches on disk.
Future work should involve more efficient patch generation and sampling, as well as improved clustering.
\end{abstract}


\maketitle

\section{Introduction}
Cancer is an important cause of death, and of all cancers, breast cancer has the highest incidence worldwide~\parencite{WHO2020}. Cancer diagnostics is based on clinical examination, medical imaging and histopathological assessment of the tumour. The latter includes analysis of specific biomarkers that often guides treatment of the patients. Most pathology laboratories are burdened by an increasing number of biopsies and more complex diagnostics~\parencite{RCP2018}. To reduce workload for pathologists, automatic assessment of tumours and biomarkers would be of value.

A natural first step in automatic tumour and biomarker analysis would be to correctly identify the lesion, thus separating the tumour from surrounding tissue. For automatic biomarker assessment, it is important to ensure that biomarker status is obtained exclusively in the invasive epithelial cancer cells. 

With the promise of deep learning-based methods in computational pathology~\parencite{Srinidhi2021}, accurate segmentation of the cancer region would be beneficial for building new classifiers and facilitate other cancer analysis methods.

In this study, we aim to develop and validate an efficient and accurate method for automatic segmentation of invasive breast cancer tumours in histopathological whole slide images (WSIs).

\subsection{Related work}

Processing histopathological WSIs is challenging due to their large size. WSIs captured at $\times 400$ magnification may be as large as $200k \times 100k$ pixels, and as such, cannot be used directly as input to convolutional neural networks (CNNs). A solution is to downsample the image to a size that is manageable for the CNN. However, this results in loss of information and is therefore often not useful for tumour segmentation. Another widely used approach is to divide the image into smaller patches~\parencite{Aresta2019}, before each patch is sent to an algorithm to produce an output. The results are then stitched to form a complete segmentation or heatmap of the entire WSI. However, the use of such a patch-wise design based on high-resolution information only, often results in edge artifacts and poor global segmentation of larger structures~\parencite{Torres2020}.

\textcite{schmitz2020} compared multi-scale convolutional autoencoder (CAE) designs, applied in a patch-wise fashion across liver tumours in WSIs. They found that the network benefited significantly from the added multi-scale information, compared to the baseline U-Net~\parencite{unet}. They also proposed non-overlapping inference to reduce runtime at the cost of reduced accuracy along patch edges. 
For handling these edge artifacts, \textcite{Torres2020} proposed a conditional random field-based, patch-wise, merging scheme.

To improve the patch-wise design, \textcite{Guo2019} developed a multi-task network for classification and semantic segmentation of breast cancer. 
They used a pretrained InceptionV3~\parencite{InceptionV3} architecture and fine-tuned it on the Camelyon16 data set~\parencite{Bejnordi2017}. Such transfer learning has the benefit of making training more efficient, as the network is not trained from scratch. Using a more complex backbone, such as InceptionV3, has the potential benefit of improved performance. However, the architecture is computationally expensive, and might therefore not be suitable for real-time applications, such as histopathological diagnostics.

Breast cancers are known for their intra- and intertumour heterogeneity, and thus their morphological appearance varies both within and between tumours. Due to intratumour heterogeneity, the patches generated from a single WSI often contain different tissue types and a varying morphological appearance.
\textcite{Quaiser2018} studied the effect of smart patch selection and balancing in preprocessing, to produce models that performed well on varying types of tissue. They demonstrated that a deep clustering approach of patches outperformed the conventional $k$-means~\parencite{kmeans1956} clustering method.

A similar cluster-guiding strategy was performed by \textcite{Yao2019} using multiple instance learning. They used a pretrained VGG19-encoder~\parencite{Simonyan2015VeryDC} for feature extraction. The dimensionality of the features was reduced using principal component analysis (PCA)~\parencite{PCA1901}, before performing $k$-means clustering. Samples were drawn from these clusters and balanced during training. The number of clusters was set to four, as they assumed that there were four main natural tissue types in the data.

\subsection{Architecture design}
\label{archs}
For image classification, MobileNetV2~\parencite{MobileNetv2} and InceptionV3~\parencite{InceptionV3} are popular baseline architectures, benchmarked on the ImageNet Large Scale Visual Recognition Challenge (ILSVRC) data set~\parencite{ILSVRC15}. Both architectures are commonly used in digital pathology~\parencite{Aresta2019,Kassani2019,Skrede2020}.

For image segmentation, various CAE architectures have been proposed. The most commonly used CAE is \textit{U-Net}~\parencite{unet}, which is a fully-convolutional, symmetric, encoder-decoder network, including skip connections at each pooling level to efficiently produce refined segmentations. Recently, multi-scale architectures have been proposed as they can potentially extract information at different magnification levels in a more efficient and controlled manner compared to traditional CAEs~\parencite{KAMNITSAS201761}.

To further improve the multi-scale design, deep supervision has been proposed~\parencite{Nie2017}. By combining multi-scale input with deep supervision, the segmentation task is guided by introducing a loss at each decoder level. By forcing the network to learn representative features at all resolution levels, one aims to produce a network that better develops an understanding of the object of interest.

In addition to multi-scale designs, attention-based CAEs have become popular~\parencite{Minaee2021}. 
%The aim of attention is to force the network to emphasize the task at hand by making it learn to filter irrelevant image features or regions.
Using attention, the network learns to filter irrelevant image features or regions, thus making it more focused on the task at hand.
Two commonly used attention designs are channel-wise and position attention. Two modern multi-scale, attention-based CAEs are \textit{AGU-Net} and \textit{DAGU-Net}~\parencite{bouget2021meningioma}, which use single and dual attention, respectively. Both architectures also use deep supervision.

Performing both detection and semantic segmentation in a single step is challenging. Often, the segmentation result is suboptimal and a post-processing method is required. Refinement networks for the CAE itself have therefore been proposed, either end-to-end~\parencite{Jha2020} or in multiple steps~\parencite{Painchaud2019}.
The end-to-end architecture, \textit{DoubleU-Net}, is a cascaded U-Net design, where the output from the first network is sent as input to the latter, concatenated with the original input. Skip connection is then performed across the two CAEs to map learned encoder features from the initial network to the latter.

\subsection{FastPathology}
We have previously developed the open-source platform FastPathology~\parencite{Pedersen2021}, for deep learning-based digital pathology. It enables deployment of deep learning solutions on WSIs through a graphical user interface (GUI). Integration of new models and pipelines does not require coding and programming skills. The software supports a wide range of multi-input/output CNN architectures, such as MobileNetV2, U-Net and YOLOv3~\parencite{Redmon2018}. Several inference engines are also supported, such as \texttt{TensorRT}~\parencite{TensorRT}, \texttt{OpenVINO}~\parencite{OpenVINO} and \texttt{TensorFlow}~\parencite{tensorflow2015}. Through the ONNX~\parencite{bai2019} format, models can be executed using \texttt{TensorRT} or \texttt{OpenVINO}, both of which are among the fastest for graphics processing unit (GPU) (NVIDIA) and central processing unit (CPU) (Intel) inference.

\subsection{Contributions}
In this paper, we present the following contributions:
\begingroup
\renewcommand\labelenumi{(\theenumi)}
\begin{enumerate}
    \itemsep=0em
    \item A novel hierarchically-balanced, random sampling scheme that extracts patches directly from the WSI.
    \item A concurrent clustering method performed during training, without the need to store temporary results, using a novel cluster-guided loss function.
    \item A refinement network which combines high-resolution information and global information, producing superior performance over single resolution approaches.
    \item The proposed pipeline and trained models are made openly available for use in \textit{FastPathology}~\parencite{Pedersen2021}.
    \item The code to reproduce the experiments is made openly available on GitHub: \url{https://github.com/andreped/H2G-Net}.
\end{enumerate}
\endgroup

\section{Materials and methods}

\subsection{Data set and annotation design}
In this study, we used 4 $\mu$m thick whole sections (n=624) from a cohort of Norwegian breast cancer patients~\parencite{Engstrom2013}, Breast Cancer Subtypes 1 (BCS-1). All tumours were previously classified into histological grade, according to the Nottingham grading system~\parencite{Elston1991pathologicalPF}. The sections were stained with hematoxylin-eosin (H\&E), scanned at $\times400$ magnification using an Olympus scanner BX61VS with VSI120-S5, and stored in the \texttt{cellSens VSI} format using JPEG2000 compression.

For each WSI, the tumour area was delineated by pathologists using QuPath~\parencite{QuPath}. To speed up and assist with the annotation work, automatic and semi-automatic approaches were tested, similarly to the approach used by \textcite{Carse2019}.
We used two different approaches for annotation (AN1 and AN2). For both annotation designs, predicted annotations were manually adjusted by the pathologists using the brush tool in QuPath.

The first 150 WSIs were annotated using the \textbf{AN1} method, which involved using the semi-automatic tissue detection function in QuPath. The following parameters were used for performing segmentation: simple tissue detection threshold 200, requested pixel size 20, and minimum area 100,000. In cases were the algorithm failed, the parameters were adjusted or the tumour was manually annotated from scratch.

\begin{figure}[h!]
    \centering
    \includegraphics[width=1\columnwidth]{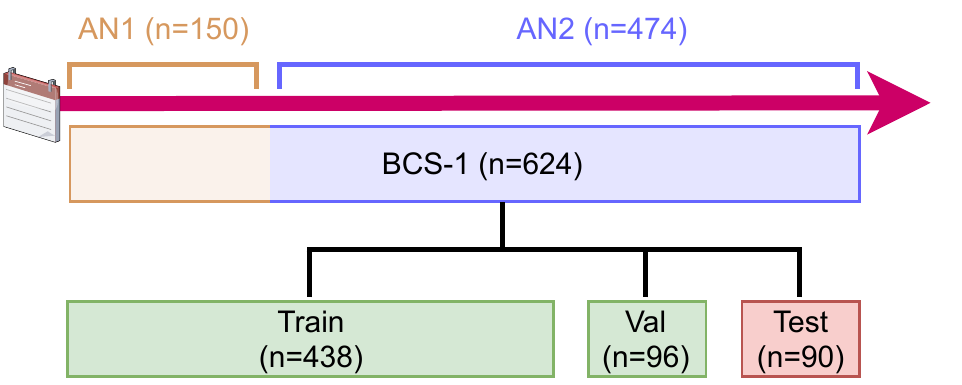}
    \caption{Description of the data generation timeline and process. The 624 WSIs were annotated with two different annotation methods (AN1 and AN2). The data set was then randomly split into train, validation, and test sets. AN: Annotation, BCS: Breast Cancer Subtypes. Val: Validation.}
    \label{fig:data-timeline}
\end{figure}

The remaining WSIs (n=474) were annotated using the \textbf{AN2} method (see \cref{fig:data-timeline}). A patch-wise CNN, similar to the Inc-PW method described in \cref{baseline_methods}, was trained from a subset of the first 150 annotated WSIs. The model was trained in Python, and the produced model was then applied to the remaining WSIs. The resulting heatmaps were imported in QuPath and converted to annotations. Simple morphological post-processing was then performed before the segmentations were adjusted by the pathologists.

Finally, to ensure consistency, all annotations were reviewed by a single pathologist experienced in breast cancer pathology and minor adjustments were made.

The pathologists' annotations were exported from QuPath as individual \texttt{PNG}s, one for each WSI, with a downsampling factor of four. The \texttt{PNG}s were then converted to tiled, pyramidal \texttt{TIFF}s, using the command line tool \texttt{vips}\footnote{\url{https://github.com/libvips/libvips}}, with tiles sized $1024\times 1024$ and a \texttt{LZW} lossless compression. All WSIs were converted to the single-file, pyramidal tiled, generic \texttt{TIFF} format using the command line tool \texttt{vsi2tif}\footnote{\url{https://github.com/andreped/vsi2tif}}.

Lastly, the annotated WSIs were randomly distributed into the three sets: training ($\sim$70\%; n=438), validation ($\sim$15\%; n=96), and test ($\sim$15\%; n=90) set.

\subsection{Preprocessing}
\label{preprocessing}
Following annotation, we extracted patches sized $256 \times 256$ at $\times 100$ magnification level from the WSIs. Only patches containing more than 25\% tissue were included. Patches with more than 25\% tumour were considered tumour patches, and only patches with no tumour were considered non-tumour. The remaining patches in the range $(0, 25]\%$ tumour tissue were discarded. For each WSI, the coordinates of accepted patches were stored along with the assigned label i.e. non-tumour/tumour.

\begin{figure*}[t!]
    \centering
    \includegraphics[width=1\textwidth]{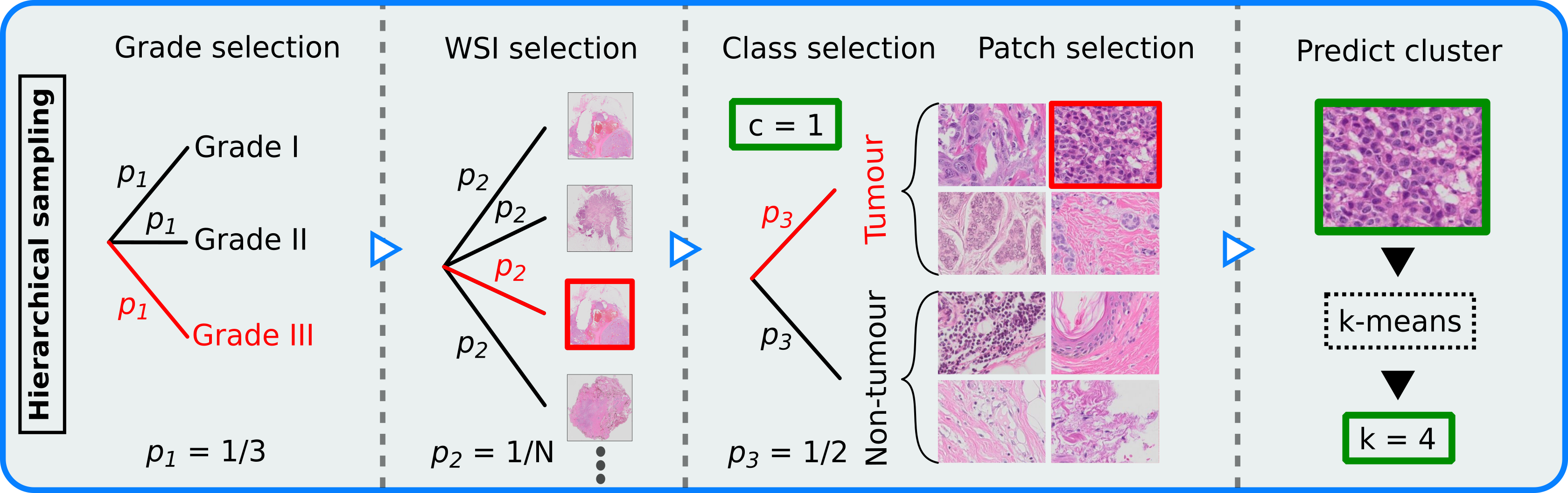}
    \caption{Illustration of the hierarchical sampling scheme, demonstrating how patches were sampled from the $N$ whole slide images (WSIs) for training the patch-wise model. Sampling was conducted as a uniform tree diagram. Thus, $p_i$ represents probability at step $i\in\{1,2,3\}$. A potential path for patch selection is marked \textcolor{red}{red}. Each patch was assigned a class label $c$ (tumour or non-tumour) and a cluster $k$ (ten different clusters). Each output is marked in \textcolor{ForestGreen}{green}.}
    \label{fig:overall_training_scheme}
\end{figure*}

\subsection{Hierarchical sampling scheme}
A batch generator was created to sample patches directly from the raw WSI format. Patches were read using OpenSlide~\parencite{Satyanarayanan2013}, which enabled multi-threading processing. The generator was based on the condition that it is important to balance patches according to the following features: class label, tissue type, tissue and tumour area, and histological grade.

Patches were sampled in a hierarchical sampling scheme (see \cref{fig:overall_training_scheme}), conducted as a tree structure uniformly distributed at each respective stage. The goal was to make all relevant outcomes equally probable. The sampling scheme was defined in the ordered stages: 1) Randomly select a histological grade, 2) from the grade select a WSI, 3) from the WSI select a class label, 4) from the class label select a patch. 
%From this selection a random patch (coordinates) was then sampled.

To include patch-level tissue type label in the balancing scheme, we used our sampling generator to train a $k$-means clustering model, similar to \textcite{Yao2019}. From a set of 100 batches of size 32, features were extracted using a VGG-16~\parencite{Simonyan2015VeryDC} backbone pretrained on the ImageNet data set~\parencite{deng2009imagenet}. The extracted features were then standardized using Z-score normalization, before PCA was performed. The number of principal components was chosen such that 95\% of the variance of the data was explained. The $k$-means model was then trained using $k=10$ number of clusters, as recommended in a related study~\parencite{Yao2020_2}. The clustering model was implemented using the Python library \texttt{scikit-learn}~\parencite{scikit-learn}.

To utilize the trained clustering model in the patch-wise CNN, \texttt{TensorFlow} equivalents of the standardization, PCA and $k$-means transform methods were implemented, which was defined as a \texttt{TensorFlow} graph. The scikit-learn trained weights were then loaded for each corresponding component. 

Finally, for training the CNN classifier, each patch was passed through two different graphs; I) a frozen pipeline that performed clustering and II) a learnable deep neural network that performed classification. The outputs from both models were then passed to the loss function.

The MobileNetV2 architecture was used for the patch-wise CNN classification of breast cancer tumour tissue, as it is lightweight, efficient, and optimized for low-end processors and thus suitable for real-time deployment. To further reduce the number of parameters, we simplified the classifier head. The updated classifier contained a global average pooling layer, followed by a dense layer of 100 hidden neurons, dropout~\parencite{Srivastava2014DropoutAS} with a 50\% drop rate, ReLU activation, batch normalization~\parencite{BatchNormalization2015}, and finally a dense layer with softmax activation function.

\begin{figure*}[t!]
    \centering
    \includegraphics[width=0.99\textwidth]{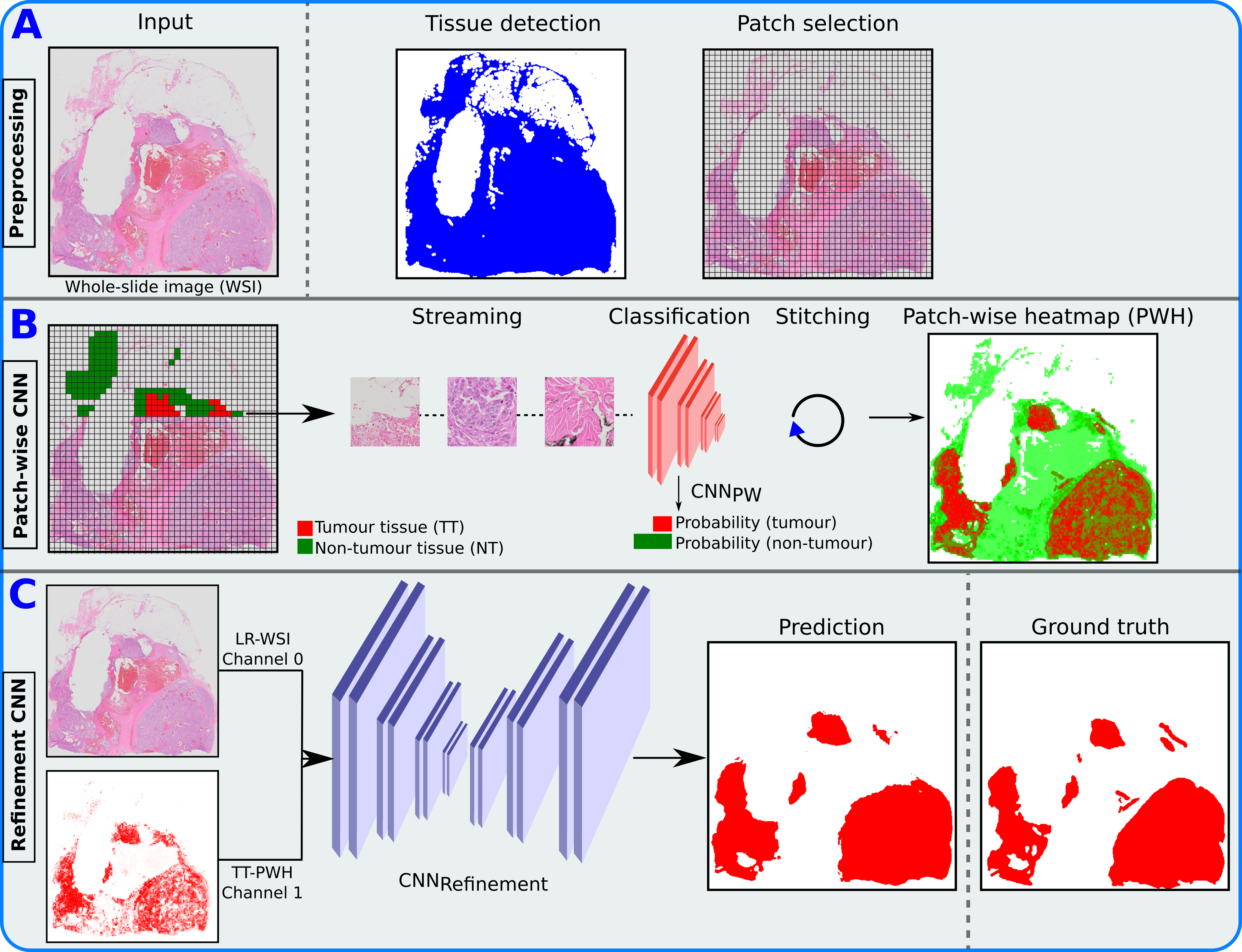}
    \caption{Illustration of the inference pipeline, from the whole slide image (WSI) to the final tumour segmentation (prediction). A) Apply tissue detection before patch selection. B) Stream accepted patches through a trained patch convolutional neural network (CNN) classifier and stitch the output to form a patch-wise heatmap (PWH). C) Merge the low-resolution (LR) WSI with the resulting tumour tissue (TT) PWH and send it through the trained refinement CNN, using a probability threshold of 0.5, to produce the final prediction.}
    \label{fig:overall-inference-scheme}
\end{figure*}

\subsection{Cluster-guided loss function}
To balance on tissue type and thus ensure similar model performance on all predicted clusters, we included the cluster-information in the loss computation. For a given batch, we calculated the cross-entropy loss for each cluster independently, and then calculated the macro average across each cluster. We named this loss function cluster-weighted categorical cross-entropy (CWCE) loss. The loss can be mathematically described as:
\begin{equation}
    \mathcal{L}_{\text{CWCE}} = -\frac{1}{K_b}\sum_{k=1}^{K_b}\sum_{c=1}^{C}\sum_{i=1}^{B}\mathds{1}(q_{i,k}=k)y_{i,c}log(p_{i,c})
\end{equation}

where $i \in \{1,...,B\}$ represents sample $i$ in a batch of size $B$, $k \in \{1,...,K_b\}$ cluster in a mini-batch $b$ of size $B$ of $K_b$ represented clusters, $c \in \{1,...,C\}$ class, $p$ class prediction, $q$ cluster prediction, and ground truth tumour class. Note that the number of clusters $K_b$ may vary between mini-batches.

\subsection{Heatmap generation}
The trained patch-wise model was then applied across the training and validation sets. Only patches assigned the tissue label were used (see  \hyperref[fig:overall-inference-scheme]{\cref*{fig:overall-inference-scheme}A}). For each WSI, the trained model was applied in a non-overlapping, sliding window fashion, storing the resulting stitched confidence heatmap of the tumour class (see \hyperref[fig:overall-inference-scheme]{\cref*{fig:overall-inference-scheme}B}).

\subsection{Refinement stage}
%Up to this point only patch-wise fined-grain information has been used for tumour detection.
To improve the result from the patch-wise detection, we combined the heatmap with the low-resolution WSI (see \hyperref[fig:overall-inference-scheme]{\cref*{fig:overall-inference-scheme}C}). 
This was done in an additional stage using a Refinement CNN.
A suitable magnification level was chosen ($\geq 1024\times 1024$ pixels), and a low-resolution version of the original WSI was extracted from the image pyramid. The image was then normalized to [0, 1], before both the resulting image and the heatmap were resized to $1024\times 1024$ using bilinear interpolation.
A separate fully-convolutional neural network was then used to refine the resulting heatmaps from the detection stage. We used the U-Net~\parencite{unet} architecture, which took the concatenated low-resolution three-channel WSI and predicted heatmap as input. 
%The low-resolution WSI was included to assist the refinement in cases where the patch-wise model performed poorly.

\subsection{From development to deployment}
After training the patch-wise and refinement models, the models are ready to be used for inference. The inference pipeline is illustrated in \cref{fig:overall-inference-scheme}.
The trained \texttt{TensorFlow} models were converted to the \texttt{ONNX} standard format, to enable efficient inference on both GPU and CPU with different frameworks. The models were then integrated into the FastPathology~\parencite{Pedersen2021} platform by writing a FAST~\parencite{Smistad2019} text pipeline, containing information about the inference pipeline and how the models should be handled (e.g., input shape, node names, and inference type). Thus, the proposed pipeline can be used through a GUI without programming.
Binary release of FastPathology, trained models, test data, and source code can be accessed on GitHub\footnote{\url{https://github.com/AICAN-Research/FAST-Pathology}}.

\section{Validation study}

\subsection{Experiments}
We conducted an ablation study to evaluate our design. The experiments conducted were:
\begin{enumerate}
    \item To assess the importance of architecture complexity in breast cancer tumour detection in WSIs, we compared CNN classifiers using the two backbone architectures InceptionV3 and MobileNetV2.
    \item To evaluate the cluster-guiding approach, we conducted experiments with and without $k$-means using the MobileNetV2 backbone.
    \item To assess the effect of post-processing on the predicted heatmap, we compared state-of-the-art CAEs against simple baseline methods.
    \item To evaluate the importance of having a GPU for inference, runtime measurements of the best performing method were performed with and without using the GPU.
\end{enumerate}

\subsection{Baseline segmentation methods}
\label{baseline_methods}
Using our pipeline, any architecture or component can be removed, added, or substituted. It is therefore valuable to assess the importance of each component in the pipeline.
To evaluate the pipeline, we used existing, well-documented, state-of-the-art architectures. For patch-wise classification we used the MobileNetV2 ($\sim 2.39$M params.) and InceptionV3 ($\sim 22.00$M params.) backbones pretrained on the ImageNet data set, and used the same simplified classifier head for both architectures, as described in \cref{archs}.

For image segmentation refinement, we compared the CAE architectures U-Net ($\sim 11.58$M params.), AGU-Net ($\sim 7.68$M params.), DAGU-Net ($\sim 9.99$M params.), and DoubleU-Net ($16.04$M params.).
In addition, we included a traditional, widely used tissue segmentation method~\parencite{Bandi2019}, to serve as a minimal baseline measure. This method simply segments all tissue, and thus all tuned methods should outperform it. For this method, the image was resized to $1024\times 1024$, before being converted to the HSV (Hue, Saturation, Value) colour domain. Then, the saturation image was thresholded using Otsu's method~\parencite{Otsu1979}.

%However, similar to what was done for the classification task,
The autoencoders were slightly modified to work better for our use case and data set. In depth details about modifications, as well as implementations for all architectures used, can be found on our GitHub repository\footnote{\url{https://github.com/andreped/H2G-Net}}.

In summary, the following segmentation designs were compared:
\begin{enumerate}[(I)]
    \itemsep=0em
    \item \textbf{Otsu}: Intensity-based thresholding for tissue segmentation.
    \item \textbf{UNet-LR}: Segmentation of low-resolution WSI using a U-Net architecture.
    \item \textbf{Inc-PW}: Patch-wise classification using an InceptionV3 architecture.
    \item \textbf{Mob-PW}: Patch-wise classification using a MobileNetV2 architecture.
    \item \textbf{Mob-KM-PW}: Same as (IV), with $k$-means guiding.
    \item \textbf{Mob-PW-UNet}: Same as (IV), with a U-Net refinement network.
    \item \textbf{Mob-PW-AGUNet}: Same as (IV), with an AGU-Net refinement network.
    \item \textbf{Mob-PW-DAGUNet}: Same as (IV), with a DAGU-Net refinement network.
    \item \textbf{Mob-PW-DoubleUNet}: Same as (IV), with a DoubleU-Net refinement network.
\end{enumerate}

\subsection{Statistical evaluation}
\label{stat-eval}
All patch-wise and refinement models were trained using the same training set, and the best models were selected based on the performance on the validation set. The test set was used as a hold-out sample for an unbiased, final evaluation.

A threshold of 0.5 was used to distinguish between the tumour and non-tumour classes. Metrics were reported WSI-wise, and only on the test set. For each respective metric, macro average and standard deviation were reported. The specific metrics used to assess performance were pixel-wise recall, precision, and the dice similarity coefficient (DSC).
To further assess the robustness of the design, we also reported DSC for each histological grade. 

We performed multiple pairwise Tukey's range tests, comparing the DSC measures for all deep learning-based designs. The p-values were estimated for the test set (see Supplementary, Table S\url{1}).

\subsection{Training parameters}
For training the classification models, we fine-tuned the respective pretrained backbones using the Adam optimizer~\parencite{adam2014} with an initial learning rate of 1e-4. For batch generation, 500 and 200 batches of size 64 for training and validation, respectively, were sampled randomly for each epoch. The models were trained for 100 epochs. Batches were generated in parallel using eight workers with a maximum queue size of 20. Models were trained using the following online data augmentation scheme of which all had a 50\% chance of being used: random horizontal/vertical flip, 90$^{\circ}$ lossless rotations, HSV colour augmentation with a random shift of range [-20, 20], and multiplicative brightness augmentation of range [0.8, 1.2].

All segmentation models were trained from scratch using the Adam optimizer with an initial learning rate of 1e-3. Accumulated gradients using a batch size of four with six accumulation steps were performed. For online data augmentation, simple horizontal/vertical flip, 90$^\circ$ rotations, random zoom of range [0.8, 1.2], and Macenko~\parencite{Macenko2009} stain augmentation\footnote{\url{https://github.com/Peter554/StainTools}} using $\sigma_1=\sigma_2=0.1$, with a chance of 50\% of being used, were conducted. The models were trained for 1000 epochs, or until the early stopping criterion with a patience of 100 epochs was achieved.

Implementation was done in Python 3.6, and CNN architectures were implemented in TensorFlow (v1.13.1). Experiments were performed using an Intel Xeon Silver 3110 CPU, with 32 cores and 2.10 GHz, and an NVIDIA Quadro P5000 dedicated GPU.

\section{Results}
For the test set, all deep learning-based methods outperformed the tissue segmentation method, Otsu, in terms of DSC (see \cref{tab:seg_all}). Comparing the patch-wise classifiers, Inc-PW and Mob-PW, no significant difference in DSC was found between the architectures ($p \approx 0.9$, see Supplementary, Table S\url{1}). Adding cluster-guiding to Mob-PW, Mob-PW-KM, slightly reduced performance, however, not significantly ($p \approx 0.9$).

\begin{table}[h!]
    \centering
    \caption{Test set segmentation performance for the different designs. DSC: Dice Similarity Coefficient, Inc: InceptionV3, Mob: MobileNetV2, PW: Patch-wise, KM: $k$-means, LR: Low-resolution. Results are reported as mean $\pm$ standard deviation.}
    \LARGE
    \adjustbox{width=\linewidth}{
    \begin{tabular}{rlccc}
        \toprule
        %\multicolumn{2}{l}{\textbf{Architecture}} & \textbf{Recall} & \textbf{Precision} & \textbf{DSC} \\
        & \textbf{Designs} & \textbf{Recall} & \textbf{Precision} & \textbf{DSC} \\
        \midrule
        \textbf{(\rom{1})} & \textbf{Otsu} & $0.990 \pm 0.027$ & $0.534 \pm 0.200$ & $0.669 \pm 0.179$ \\
        \textbf{(\rom{2})} & \textbf{UNet-LR} & $0.931 \pm 0.113$ & $0.851 \pm 0.165$ & $0.874 \pm 0.128$ \\
        \textbf{(\rom{3})} & \textbf{Inc-PW} & $0.881 \pm 0.118$ & $0.909 \pm 0.099$ & $0.887 \pm 0.089$ \\
        \textbf{(\rom{4})} & \textbf{Mob-PW} & $0.879 \pm 0.123$ & $0.907 \pm 0.100$ & $0.885 \pm 0.094$ \\
        \textbf{(\rom{5})} & \textbf{Mob-KM-PW} & $0.853 \pm 0.124$ & $0.909 \pm 0.097$ & $0.872 \pm 0.092$ \\
        \textbf{(\rom{6})} & \textbf{Mob-PW-UNet} & $0.944 \pm 0.074$ & $\boldsymbol{0.929 \pm 0.088}$ & $\boldsymbol{0.933 \pm 0.069}$ \\
        \textbf{(\rom{7})} & \textbf{Mob-PW-AGUNet} & $\boldsymbol{0.954 \pm 0.066}$ & $0.909 \pm 0.097$ & $0.927 \pm 0.072$ \\
        \textbf{(\rom{8})} & \textbf{Mob-PW-DAGUNet} & $0.942 \pm 0.075$ & $0.922 \pm 0.091$ & $0.928 \pm 0.072$ \\
        \textbf{(\rom{9})} & \textbf{Mob-PW-DoubleUNet} & $0.949 \pm 0.073$ & $0.919 \pm 0.093$ & $0.929 \pm 0.074$ \\
        \bottomrule
    \end{tabular}
    }
    \label{tab:seg_all}
\end{table}

Among the best single-resolution designs (i.e. UNet-LR, Inc-PW, and Mob-PW), the patch-wise approaches performed slightly better in terms of DSC, but not significantly ($p \approx$ 0.9). The low-resolution approach (UNet-LR) achieved better recall, but with the cost of poorer precision.

Introducing a U-Net-inspired refinement network (using both the low-resolution WSI and the resulting heatmap from Mob-PW) resulted in significant improvement compared to the best single resolution approach (Mob-PW-UNet vs Inc-PW, $p \approx 0.012$). All methods using a refinement network significantly outperformed the single resolution approaches.

\begin{table}[h!]
    \centering
    \caption{Test set segmentation performance for the different designs in histological grades I-III. DSC: Dice Similarity Coefficient, Inc: InceptionV3, Mob: MobileNetV2, PW: Patch-wise, KM: $k$-means, LR: Low-resolution. Results are reported as mean $\pm$ standard deviation.}
    \LARGE
    \adjustbox{width=\linewidth}{
    \begin{tabular}{rlccc}
        \toprule
        %\multicolumn{2}{l}{\multirow{2}{*}{\textbf{Architecture}}} & \multicolumn{3}{c}{\textbf{DSC} (n=90)} \\\cmidrule(lr){3-5}
        & \multirow{2}{*}{\textbf{Designs}} & \multicolumn{3}{c}{\textbf{DSC} (n=90)} \\\cmidrule(lr){3-5}
        & & \textbf{Grade $\boldsymbol{\mathrm{I}}$ (11)} & \textbf{Grade $\boldsymbol{\mathrm{II}}$ (48)} & \textbf{Grade $\boldsymbol{\mathrm{III}}$ (31)} \\
        \midrule
        \textbf{(\rom{1})} & \textbf{Otsu} & $0.732 \pm 0.151$ & $0.659 \pm 0.186$ & $0.664 \pm 0.174$ \\
        \textbf{(\rom{2})} & \textbf{UNet-LR} & $0.880 \pm 0.127$ & $0.862 \pm 0.142$ & $0.890 \pm 0.099$ \\
        \textbf{(\rom{3})} & \textbf{Inc-PW} & $0.901 \pm 0.072$ & $0.882 \pm 0.088$ & $0.890 \pm 0.095$ \\
        \textbf{(\rom{4})} & \textbf{Mob-PW} & $0.887 \pm 0.089$ & $0.882 \pm 0.092$ & $0.890 \pm 0.100$ \\
        \textbf{(\rom{5})} & \textbf{Mob-KM-PW} & $0.851 \pm 0.111$ & $0.872 \pm 0.089$ & $0.880 \pm 0.088$ \\
        \textbf{(\rom{6})} & \textbf{Mob-PW-UNet} & $0.936 \pm 0.073$ & $\boldsymbol{0.931 \pm 0.058}$ & $\boldsymbol{0.935 \pm 0.083}$ \\
        \textbf{(\rom{7})} & \textbf{Mob-PW-AGUNet} & $0.933 \pm 0.082$ & $0.926 \pm 0.060$ & $0.927 \pm 0.083$ \\
        \textbf{(\rom{8})} & \textbf{Mob-PW-DAGUNet} & $0.935 \pm 0.075$ & $0.926 \pm 0.058$ & $0.929 \pm 0.088$ \\
        \textbf{(\rom{9})} & \textbf{Mob-PW-DoubleUNet} & $\boldsymbol{0.942} \pm 0.070$ & $0.924 \pm 0.066$ & $0.934 \pm 0.085$ \\
        \bottomrule
    \end{tabular}
    }
    \label{tab:seg_grade}
\end{table}

Comparing the refinement architectures, the best performance in terms of precision and DSC was found from the U-Net design, Mob-PW-UNet, but the difference was not statistically significant ($p \approx 0.9$ for all comparisons). No benefit of using more advanced CAE architectures was found.

When each histological grade was analyzed separately, Mob-PW-DoubleUNet was the best performing method for grade \rom{1} and Mob-PW-UNet performed best on grade \rom{2} and \rom{3} (see \cref{tab:seg_grade}). All designs guided by the hierarchical sampling scheme (designs (II)-(IX)) performed similarly across all histological grades, indicating that performance was somewhat invariant to histological grade.

\begin{table}[h!]
    \centering
    \caption{Runtime measurements of the proposed method, Mob-PW-UNet. Experiments were repeated ten times and respective average and standard deviation are reported in seconds. OpenVINO and TensorRT were used as inference engines for CPU and GPU inference, respectively.}
    \resizebox{6.5cm}{!}{
    \begin{tabular}{llccc}
        \toprule
        \textbf{} & \textbf{Patch-wise} & \textbf{Refinement} & \textbf{Total} \\
         \midrule
         \textbf{OpenVINO} & $57.32 \pm 0.20$ & $0.75 \pm 0.01$ & $58.07 \pm 0.20$ \\
        \textbf{TensorRT} & $39.88 \pm 0.62$ & $0.38 \pm 0.00$ & $40.26 \pm 0.62$ \\
    \end{tabular}
    }
    \label{tab:runtime}
\end{table}

The best performing method, Mob-PW-UNet, took approximately $\sim58$ seconds to run on a representative $\times$400 WSI using the CPU (see \cref{tab:runtime}). Using the GPU, runtime was reduced to 40.26 seconds. The patch-wise method dominated the overall runtime, with $\sim1$\% of the total runtime being used on the refinement stage.

\section{Discussion}
%% start by mentioning what has been done in this study
In this paper, we have developed a pipeline for tumour segmentation of WSIs, called H2G-Net, using H\&E-stained WSIs from a well-described cohort of breast cancer patients. We have presented each component in the pipeline and assessed the impact of each component in an ablation study. Using multiple guiding components, we significantly improved segmentation performance, while reducing disk storage requirements compared to traditional training pipelines.

The best performing architectures utilized both low and high-resolution information from the WSI. A similar approach is used by pathologists when separating the tumour from surrounding tissue. The low-resolution image provides a coarse outline of the tumour, whereas higher resolution is often necessary for accurate delineation.

\subsection{Cascaded design and related work}
For segmentation, using low-resolution as the first step could reduce total runtime by filtering patches during preprocessing. However, UNet-LR, a U-Net using only low-resolution information, results in low sensitivity and should therefore not be the first step for breast cancer segmentation. In this work, we used a patch-wise, high-resolution method as a first step to optimize detection. A U-Net could then be trained at a later stage to refine the produced heatmap, using both the heatmap and the low-resolution WSI as input.
Using the heatmap generated from the patch-wise method alone, some areas of the tumour, such as areas with abundant stromal or fatty tissue, may not be recognized.
We show that the network benefits from having the low-resolution image, together with the heatmap.

% comparing to related work  In contrast to our refinement design, 
\textcite{Tang2021} used a two-step procedure to perform instance segmentation of objects in the Cityscapes data set~\parencite{Cordts2016Cityscapes}. These images have initial resolution of $1024 \times 2048$ pixels, but the classes of interest can easily be distinguished at lower resolution. In contrast to our design, they first performed semantic segmentation on the low-resolution, before refining the initial segmentation using a patch-wise design. 
They also used overlapping predictions along the border of the initial segmentation. Their approach might be an interesting refinement method to further improve the segmentation performance of our resulting low-resolution segmentation.

A similar refinement approach to \textcite{Tang2021} was used by \textcite{Sornapudi2020} for cervical intraepithelial neoplasia segmentation of H\&E stained WSIs. However, introducing a new network for border refinement will make the overall runtime longer and introduce more complexity to the final pipeline.
\textcite{Isensee2021} conducted a similar two-step procedure for medical volumetric data. They first used a CAE applied on the downsampled version of the full 3D volume (CT/MRI). They then applied a 3D patch-wise refinement model, using both the local volumetric data (CT/MRI) and predicted heatmap as input.

Another similar architecture design to ours, was proposed by \textcite{Nazeri2018} for WSI classification of breast cancer. They also used a patch-wise model in the first step, before feeding the resulting heatmap to a second CNN that performed WSI-level classification. In addition, they used skip connections to propagate learned features from the patch-wise CNN to the latter CNN. Thus, our method could be seen as an adaption to their design applied to image segmentation. This style of skip connection is similar to the Double U-net approach by \textcite{Jha2020}. In this study, we did not explore skipping features from the classifier to the refinement network. This could be explored in future work.
Our design is also similar to the work of \textcite{Daly2020}, where a similar two-stage, cascaded CNN design was deployed, but for image registration of WSIs.

\subsection{Architecture depth, clustering, and patch generation}
An interesting observation in this study is that using the deepest and most complex network for tumour segmentation is not necessarily better. From \cref{tab:seg_all}, we observe that InceptionV3 outperformed MobileNetV2 without any of the proposed guiding methods. However, we can achieve similar accuracy to InceptionV3, in addition to reduced inference memory usage and runtime, by using cluster-guiding with the MobileNetV2 architecture.
A similar trend could be seen from the refinement network. Choosing more complex CAEs did not significantly improve segmentation performance. This could be due to data that did not cover all possible variations. The quality of the heatmap provided from the detection stage varied in some cases, making it challenging for the refinement network to improve the initial segmentation.

Reading patches from the raw WSI format is time consuming. It is therefore common to preprocess data before training.
In this study, we sampled patches directly from the raw WSI format during training. This idea was recently proposed by \textcite{lutnick2021histofetch}. We further extended on their idea to make it more generic.
The approach by \textcite{lutnick2021histofetch} cannot handle larger batch sizes, as the cost of batch generation is not scalable. Thus, we used accumulated gradients to speed up batch generation, while simultaneously reducing GPU memory usage. We further introduced the concept of hierarchical sampling, which added direct support for balancing on multiple categories and labels. This design also added direct support for performing cluster-balancing end-to-end during training. Even though our design did not benefit from the clustering method used, the clustering method could easily be substituted.

\subsection{Future perspectives}

No benefit from using cluster-guiding to detect breast cancer tissue was observed, comparing Mob-KM-PW and Mob-PW by qualitative visual inspection. This was also observed in a study by \textcite{Quaiser2018}. They demonstrated improved performance by using a more advanced clustering approach. Thus, in future work, substituting the ImageNet features + PCA + $k$-means clustering approach with a more suited clustering design should be explored. Given a suitable clustering approach, a natural next step could be to provide the predicted cluster heatmap with the confidence map, as it would provide different, representation-informative, high-resolution information to the refinement network.

% Multiple Instance Learning
Multiple instance learning (MIL) is a promising approach that tackles the challenge of weak supervision and noisy ground truth~\parencite{Ilse2018}. Exchanging the MIL design with the single-instance CNN classifier is possible. In this framework, one could still perform clustering in preprocessing, and sample patches to the bag, as done in a recent study~\parencite{Yao2020_2}. However, an interesting approach proposed by \textcite{Xu2012} was to perform clustering directly within the MIL design on bag-level. Using attention, one could train a network, not only to solve a task, but to learn subcategory structures in the data, while simultaneously filtering redundant clusters and noisy patches. This was demonstrated by \textcite{Lu2020}. However, they did not assess the impact of the clustering component. Future work should involve replacing the single-instance CNN with the MIL design, incorporating clustering in an end-to-end fashion, and properly assessing its impact.

\subsection{Strengths and limitations}
The main strengths of the study are that the models were trained on a large set of breast cancer WSIs.
Tumour annotations were created in a (semi-)automatic manner, and manually corrected by pathologists.
To ensure consistency, all annotations were assessed by a pathologist experienced in breast cancer pathology.
Validation studies were conducted using an independent test set.
The performance of the different designs was also evaluated for each histological grade separately. An ablation study was performed to assess the impact of each component in the multi-step pipeline. The proposed design was validated against baseline methods, and the best method was integrated in an open platform, FastPathology~\parencite{Pedersen2021}.

The main limitations are that the models were trained using sections that were H\&E-stained in the same laboratory and scanned in a single scanner. We demonstrated that the model generalized well to the test set, however, we have not tested our model on WSIs from other institutions. It is possible to carry out data augmentation to make the models more invariant, but it is challenging to mimic different staining and scanning effects~\parencite{TELLEZ2019101544}. Thus, in the future, data from different laboratories and scanners will be added for training the models. 
We did not perform stain normalization as it would have added an additional layer of uncertainty and dependency in the pipeline.
Furthermore, it would be interesting to assess the extent of generalization capability of our models to cancers of other origins, such as lung or gastrointestinal cancer.

\section{Conclusion}
Through our hybrid guiding scheme, we demonstrated a significant improvement in segmentation of breast cancer tumours from gigapixel histopathology images. The model outperformed single resolution approaches, and introduced a simple, fast, and accurate way to refine segmentation heatmaps, without the need for overlapping inference or ensembling.
We also presented a hierarchical sampling scheme, that enabled patches to be streamed from the raw WSI format concurrently during training. Furthermore, we demonstrated that tissue type balancing can be performed end-to-end, using a novel loss function. The hierarchical sampling scheme and the novel loss function were introduced to make training methods more scalable and to reduce storage requirements.

\section*{Authors' contribution}
Andr\'e Pedersen: Conceptualization, Methodology, Investigation, Annotation, Writing – original draft, Writing - review \& editing, Software, Validation. Erik Smistad: Methodology, Investigation, Supervision, Writing – original draft, Writing - review \& editing, Software. Tor V. Rise: Annotation, Writing - review \& editing, Pathology expertise. Vibeke G. Dale: Annotation, Writing - review \& editing, Pathology expertise. Henrik S. Pettersen: Annotation, Writing - review \& editing, Pathology expertise. David Bouget: Methodology, Supervision, Writing - original draft, Writing - review \& editing. Tor-Arne S. Nordmo: Methodology, Investigation, Writing - review \& editing. Ingerid Reinertsen: Conceptualization, Methodology, Supervision, Writing - original draft, Writing - review \& editing. Marit Valla: Conceptualization, Methodology, Investigation, Annotation, Supervision, Data curation, Writing - original draft, Writing - review \& editing, Resources, Pathology expertise, Project administration. All authors reviewed and approved the final version of the manuscript.

\section*{Funding}
This work was supported by The Liaison Committee for Education, Research and Innovation in Central Norway [grant number 2018/42794]; The Joint Research Committee between St. Olavs hospital and the Faculty of Medicine and Health Sciences, NTNU (FFU) [grant number 2019/38882]; The Cancer Foundation, St. Olavs hospital, Trondheim University Hospital [grant number 13/2021]; and The Clinic of Laboratory Medicine, St. Olavs hospital, Trondheim University Hospital [grant number 2020/14728-49].

\section*{Conflicts of interest statement}
The authors declare no conflict of interest.

\printbibliography

\end{document}